\documentclass[aps,preprint,longbibliography]{revtex4-1}

\usepackage{amsmath}
\usepackage[caption=false]{subfig} 
\usepackage{graphicx}
\usepackage[colorlinks=true,allcolors=blue]{hyperref}
\usepackage{soul}

\usepackage{todonotes}

\def\lsim{\mathrel{\rlap{\lower4pt\hbox{\hskip1pt$\sim$}}
    \raise1pt\hbox{$<$}}}
\def\gsim{\mathrel{\rlap{\lower4pt\hbox{\hskip1pt$\sim$}}
    \raise1pt\hbox{$>$}}}

\begin{document}
\preprint{MIT-CTP/5941}

\title{Transport properties of stochastic fluids}

\author{Chandrodoy Chattopadhyay$^1$, 
Josh Ott$^2$, 
Thomas Sch\"afer$^3$,  
and Vladimir V. Skokov$^{3}$}

\affiliation{$^1$ Theoretical Physics Division, Physical
Research Laboratory, Navrangpura, Ahmedabad 380009, India}
\affiliation{$^2$ Center for Theoretical Physics 
-- a Leinweber Institute, Massachusetts Institute 
of Technology, Cambridge, MA 02139}
\affiliation{$^3$ Department of Physics, 
North Carolina State University,
Raleigh, NC 27695}
\begin{abstract}
We study heat conduction and momentum transport in the context
of stochastic fluid dynamics. We consider a fluid described 
by model H in the classification of Hohenberg and Halperin. 
We study both non-critical and critical fluids, and we investigate
transport properties in two as well as three dimensions. Our
results are based on numerical simulations of model H using 
a Metropolis algorithm, and we employ Kubo relations to extract transport coefficients. We observe the expected logarithmic
divergence of the shear viscosity in a two-dimensional
non-critical fluid. At a critical point, we find that the 
transport coefficients exhibit power-law scaling with the 
system size $L$. The strongest divergence is seen for the 
thermal conductivity $\kappa$ in two dimensions. We find
$\kappa\sim L^{x_\kappa}$ with $x_\kappa=1.6\pm 0.1$. The 
divergence is weaker in three dimensions, $x_\kappa=1.25
\pm 0.3$, and the scaling exponent for the shear viscosity,
$x_\eta$, is significantly smaller than $x_\kappa$ in both
two and three dimensions. 
\end{abstract}

\maketitle

\section{Introduction}
\label{sec:intro}

 In recent years there has been significant interest in the 
 role of fluctuations in fluid dynamics \cite{Martin:1973,Pomeau:1975,
Hohenberg:1977ym,Kovtun:2012rj,Crossley:2015evo,Huang:2023eyz,
Basar:2024srd}. Stochastic effects are important for a number of
reasons. First of all, fluctuations and the fluctuation-dissipation 
relation are a central ingredient in formulating fluid dynamics as 
an effective field theory \cite{Martin:1973,Crossley:2015evo,
Huang:2023eyz}. Second, fluctuations lead to new phenomena, including 
non-analytic long-time tails, that may be phenomenologically important, 
but that are missed by the naive gradient expansion \cite{Pomeau:1975,
Kovtun:2012rj}. Finally, the dynamics of fluctuations is essential 
to any description of a fluid in the vicinity of a phase transition
\cite{Hohenberg:1977ym,Basar:2024srd,Bzdak:2019pkr,Bluhm:2020mpc}.
In particular, the dynamics of fluctuations accounts for critical
slowing down, dynamic scaling, and critical transport behavior
\cite{Hohenberg:1977ym}.

  In a non-critical fluid fluctuation corrections are typically 
perturbative, and they can be studied systematically using a variety 
of methods, including effective actions and kinetic equations
\cite{Kovtun:2012rj,Akamatsu:2016llw}. While the effective action 
approach has only been studied in homogeneous and static backgrounds, 
the hydro-kinetic method has also been extended to more complicated 
backgrounds. In particular, Akamatsu et al.~\cite{Akamatsu:2018vjr,
Akamatsu:2018vjr} studied the dynamics of fluctuations in a 
Bjorken background, and An and collaborators \cite{An:2020vri,
An:2022jgc} constructed a set of deterministic equations for 
the evolution of equal time $n$-point functions of hydrodynamic 
variables on a general background. 

  Historically, critical dynamics has mainly been studied using 
the epsilon expansion \cite{Siggia:1976,Hohenberg:1977ym}, or  
based on non-systematic approximations known as ``mode-coupling"
theories \cite{Kawasaki:1970,Ohta:1976}. More recently, a number 
of authors have employed the functional renormalization group 
in order to extract dynamical exponents and spectral functions
\cite{Canet:2006xu,Canet:2011wf,Mesterhazy:2013naa,Chen:2023tqc,
Roth:2023wbp,Roth:2024hcu}. These methods have been very successful, 
but they do involve assumptions and approximations, and they are
not easily generalized to non-trivial backgrounds. 

 In order to address these issues a number of groups have studied 
numerical simulations of stochastic fluid dynamics \cite{Bell:2007,
Donev:2010,Berges:2009jz,Schweitzer:2020noq,Nahrgang:2018afz,
Nakano:2025}. Until recently, these studies have typically been 
restricted to the linearized theory \cite{Bell:2007,Donev:2010}. 
Significant progress was made using a Metropolis algorithm. This 
algorithm combines the diffusive and stochastic updates into a single 
Metropolis step, and ensures that fluctuation-dissipation relations 
are satisfied. The method has been used to study relaxational dynamics
(model A) \cite{Schaefer:2022bfm}, diffusive dynamics (model B) 
\cite{Chattopadhyay:2023jfm}, chiral dynamics (model G) 
\cite{Florio:2021jlx,Florio:2023kmy}, and the diffusive dynamics
of a conserved order parameter coupled to the momentum density of the
fluid (model H) \cite{Chattopadhyay:2024jlh,Chattopadhyay:2024bcv}. 
Here, we follow the nomenclature of Hohenberg and Halperin, who
provided a classification of hydrodynamic theories of critical 
dynamics \cite{Hohenberg:1977ym}. Model H is of particular interest, 
as it is expected to describe the dynamics in the vicinity of a possible
endpoint in the QCD phase diagram \cite{Son:2004iv}.

 In our previous work on model H we considered a static 
background in a box of volume $(La)^3$ both at and away 
from the critical point and performed a number of checks
\cite{Chattopadhyay:2024jlh,Chattopadhyay:2024bcv}. We 
studied the renormalization of the shear viscosity away 
from the critical point, known as the ``stickiness of 
sound (or shear)'' \cite{Kovtun:2011np}. We verified 
dynamic scaling near the critical point, and measured 
the dynamic scaling exponent $z$ in two and three-dimensional
fluids.

  In the present work we wish to go further and study transport 
properties of critical and non-critical fluids. We explore the 
logarithmic divergence of the shear viscosity in non-critical 
two-dimensional fluids, and the critical divergence of both 
the shear viscosity and the thermal conductivity in two and 
three dimensions. We will examine the critical dynamics in 
model H, as well as in a truncation in which the self-adevection
of the momentum density is neglected. We refer to this truncation
as model H0 \footnote{Different names are used in the 
literature. Indeed, Hohenberg and Halperin use ``model H'' for the 
truncation we call model H0.}.
We have several goals in mind. First, we want 
to provide additional consistency checks for our methods. Second, 
we wish to obtain numerical values for the critical exponents 
associated with transport properties that are independent of 
additional assumptions or truncations. Finally, we want to 
provide a baseline for studies of the importance of critical
transport behavior in expanding fluids. 

  The manuscript is structured as follows: In Sect.~\ref{sec:transport}
we quickly review the equations of motion of model H, as well as
different approaches to measuring transport properties. We focus, 
in particular, on Kubo relations. In Sect.~\ref{sec:theory} we 
summarize theoretical expectations in the non-critical and critical 
regimes. Numerical results are described in Sect.~\ref{sec:num}, 
and a brief outlook is provided in Sect.~\ref{sec:sum}.

\section{Critical and Non-critical transport}
\label{sec:transport}

\subsection{Model H}
\label{sec:modH}

 Model H is a theory that describes the coupling of a
scalar order parameter field $\phi$ to the momentum density
$\vec{\pi}$ of the fluid. We work in a truncation where 
only the transverse components of $\vec{\pi}$ are retained, 
$\vec\nabla\cdot\vec{\pi}^T=0$. This corresponds to retaining
the effect of shear waves and neglecting the role of sound
waves. For a brief discussion of the role of sound waves 
see Sect.~II.B in \cite{Chattopadhyay:2024bcv}. The Model H 
equations can be written in conservative form \cite{Hohenberg:1977ym,
Onuki:1997,Chattopadhyay:2024jlh,
Chattopadhyay:2024bcv},
\begin{align}
\partial_t \phi  &= - \vec{\nabla} \cdot \vec{j}  \, , 
\label{modH_phi}\\
\partial_t \pi^T_i & =- P^T_{ij} \nabla_k \, \Pi_{jk}  \, .    
\label{modH_pi}
\end{align}
Here, $P_{ij}^T=\delta_{ij}-\nabla_i\nabla_j/\nabla^2$ is a
transverse projector. The fluxes are given by
\begin{align}
j_i &= \frac{1}{\rho} \, \phi \, \pi^T_i 
  - \kappa \, \nabla_i \frac{\delta {\cal H}}{\delta \phi} 
  + \theta_i \, ,  
\label{flux_phi} \\
\Pi_{ij} &= \frac{1}{\rho} \, \pi^T_i \, \pi^T_j 
   + \nabla_i \phi \nabla_j \phi 
   - \frac{\eta}{\rho} \left( \nabla_i \pi^T_j + \nabla_j \pi^T_i \right) 
   + \Lambda_{ij}\, . 
\label{flux_piT}
\end{align}
The effective hydrodynamic Hamiltonian is given by 
\begin{align}
\label{H_Ising}
    {\cal H}  = \int d^dx \left[ 
      \frac{1}{2\rho} ( \pi_i^T)^2
    +  \frac{1}{2} (\nabla \phi)^2  
    +  \frac{1}{2} m^2 \phi^2
    +   \frac{1}{4} \lambda  \phi^4  \right] \, ,
\end{align}
where $\rho$ is the mass density, $m$ is a mass parameter, and 
$\lambda$ is the non-linear self coupling. There is a critical 
value of $m^2=m_c^2$ at which the physical correlation length 
diverges and the effective Hamiltonian describes a critical 
point in the universality class of the Ising model. 
Also, $\theta_i$ 
and $\Lambda_{ij}$ are delta correlated noise terms. The correlation 
functions of the noise are given by  
\begin{align}
    \langle \theta_i (t, \vec{x}) \theta_j (t', \vec{x}') \rangle 
 &= 2 \kappa T\, \delta_{ij} \,
  \delta(\vec{x}-\vec{x}')\delta(t-t')\, ,
\label{noise-phi}  \\  
  \langle \Lambda_{ij} (t, \vec{x}) 
           \Lambda_{kl} (t', \vec{x}') \rangle &= 
    2 \eta T \left(\delta_{ik}\delta_{jl}
       + \delta_{il}\delta_{jk}-\frac{2}{3}\delta_{ij}\delta_{kl}
       \right)
      \delta(\vec{x}-\vec{x}')\delta(t-t')\, . 
\label{noise-pi} 
\end{align}
In our previous work \cite{Chattopadhyay:2024jlh,
Chattopadhyay:2024bcv} we provided a discretization of the 
model H equations on a spatial lattice with spacing $a$ and 
size $l=La$. We solved the equation of motion using a Metropolis
step that combines the diffusive and stochastic terms, and an 
approximately conserving deterministic update for the advection
term. We also introduced a truncation, called model H0, in which
the $\pi_i^T\pi_j^T$ term in $\Pi_{ij}$ is neglected. In terms 
of the equations of motion this implies that the momentum 
density $\pi_i^T$ is advected by the order parameter $\phi$,
but not by itself.

\subsection{Measuring viscosity and thermal conductivity}
\label{sec:visc}

 We have previously studied the shear viscosity of a stochastic
fluid using the statistical correlation function of the momentum
density \cite{Chattopadhyay:2024bcv}
\begin{align}
 C_{ij}(t,\vec{k}) = \left\langle \pi_i^T(0,\vec{k})
    \pi_j^T(t,-\vec{k}) \right\rangle \, .
\label{C-pi}
\end{align}  
Here, $\pi^T_i(t,\vec{k})$ is the Fourier transform of 
$\pi^T_i(t,\vec{x})$ with respect to the spatial variable 
$\vec{x}$. The transversality of $\pi_i^T$ implies that 
$C_{ij}(t,\vec{k})= (\delta_{ij}-\hat{k}_i\hat{k}_j)
C_\pi(t,k)$. In linearized hydrodynamics this correlation
function is governed by the shear mode and 
\begin{align}
    C_\pi(t,k) = \rho T\, \exp\left(-(\eta/\rho)k^2t\right)\, .
\label{C-pi-lin}
\end{align}
This formula implies that the momentum diffusivity $D_\eta=
\eta/\rho$ can be extracted from the logarithmic derivative 
of the correlation function. Note that the momentum diffusivity
can be written as $D_\eta=\eta\chi^{-1}_\pi$, where $\chi_\pi
=\rho$ is the momentum susceptibility. Gallilean invariance 
fixes the functional form of the stress tensor $\Pi_{ij}=(\pi_i^T
\pi_j^T)/\rho$, and the susceptibility does not get renormalized.
As a consequence, the exponential decay of $C_\pi$ can be used 
to measure the viscosity $\eta$. 

 This measurement relies on a number of assumptions. First, the wave 
number has to be  sufficiently small so that the decay is controlled 
by viscosity. On dimensional grounds we expect higher order terms to 
be suppressed if $k^2\lsim  T/D_\eta $. Furthermore, we need to assume
that even though the diffusion constant $D_\eta$ is modified by 
fluctuations, the exponential decay $C_\pi(t,k)\sim \exp(-D_\eta k^2t)$ 
is not. This assumption is not correct. In perturbation theory shear 
waves with wave number $k$ can decay into modes with wave number
$k/2$. This leads to a diffusive cascade, which generates 
contributions of the form $C_\pi \sim n!/t^{3n/2}\exp(-D_\eta k^2t
/n)$, where $n$ is the number of diffusive modes emitted in the 
cascade. Delacretaz conjectured that this cascade is resummed into 
a modified exponential decay $C_\pi(t,k)\sim \exp(-\alpha\sqrt{ 
D_\eta k^2 t})$ where $\alpha$ is a numerical coefficient 
\cite{Delacretaz:2020nit}.

 In $d=3$ we have not observed deviations from the simple
exponential decay behavior $C_\pi(t,k)\sim \exp(-D_\eta k^2 t)$, 
see Fig.~6 in Ref.~\cite{Chattopadhyay:2024bcv}. This is 
likely related to the fact that we consider the lowest wave
number on a finite lattice. This mode can get renormalized by 
the coupling to high momentum modes, but it cannot initiate a
diffusive cascade towards lower momenta. Indeed, we observe a
renormalization of the viscosity, but no modification of the 
long-time behavior. Near a critical point we expect the viscosity 
to scale with the correlation length, $\eta\sim \xi^{x_\eta}$. 
In a finite system at the critical point the correlation length 
is limited by the system size, and the scaling relation is 
$\eta\sim L^{x_\eta}$. 

 We can apply the same idea to the measurement of the thermal
conductivity. The correlation function of the order parameter 
is 
\begin{align}
C_\phi(t,\vec{k}) = \langle \phi(0,\vec{k})
      \phi(t,-\vec{k})\rangle .
\end{align}
In the hydrodynamic regime $C_\phi(t,k)\sim \exp(-D_\kappa k^2t)$
where $D_\kappa=\kappa\chi_\phi^{-1}$ and $\chi_\phi$ is the order 
parameter susceptibility. The exponential decay determines the 
diffusion constant, but a determination of $\kappa$ requires
an independent measurement of the susceptibility. 

In the non-critical regime we have the same considerations mentioned 
before in connection with the momentum density. The wave number $k$ 
has to be in the hydrodynamic regime, and in general we expect the 
appearance of long-time tails that modify the exponential decay. 
An additional issue arises in the critical regime.  
We will see in Sect.~\ref{sec:theory-crit} that diffusive behavior 
$C_\phi\sim \exp(-D_\kappa k^2t)$ requires $k\xi\ll 1$. On a finite 
lattice the lowest mode has wave number $k_{\it min}=\pi/(La)$, and 
$D_\kappa$ cannot be extracted at the critical point where $\xi\simeq La$. 
A measurement of the critical thermal conductivity then requires 
detuning from the critical regime, and measuring the correlation 
function $C_\phi(t,k)$ in the regime $k\xi\ll 1$ with $a\ll\xi\ll La$.

\subsection{Kubo relations}
\label{sec:Kubo}

 The discussion in the previous section indicates that there 
are a number of difficulties with extracting transport coefficients
from the asymptotic decay of two-point functions at finite wave
number. We will therefore consider measuring transport coefficients
using Kubo relations. In order to understand the issues involved
we briefly review the assumptions underlying the Kubo approach
\cite{Kubo:1957,Kadanoff:1963,Luttinger:1964zz,Kubo:1966,
Forster:1995,Mazenko:2006,Cugliandolo:2025}. 

 Consider a microscopic Hamiltonian $\hat{H}_0$, two observables 
$\hat{A}$ and $\hat{B}$, as well as an external field $h(t)$ that 
couples $\hat{B}$ to the Hamiltonian $\hat{H}(t)=\hat{H}_0-
h(t)\hat{B}$. We can study the response of $\langle \hat{A}(t)
\rangle$ to $h(t)$ in the weak field (linear response) limit 
$h\to 0$. Here, $\langle \hat{A}(t)\rangle ={\it Tr}[\hat{A}(t)
\hat\rho]$, where $\hat\rho$ is the density matrix. We define
the response function
\begin{align}
\label{resp_fct}
\chi^{AB}(t-t') = 
 \left.  \frac{\delta\langle\hat{A}(t)\rangle}{\delta h(t')}
     \right|_{h=0}   \, . 
\end{align}
The Kubo relation connects the response function to
the retarded correlation function of $\hat{A}$ and $\hat{B}$,
\begin{align}
\label{Kubo_1}
 \chi^{AB}(t-t') = 
     i\langle [\hat{A}(t),\hat{B}(t')]\rangle
     \, \theta(t-t')
     \equiv G_R^{AB}(t-t') \, . 
\end{align}
This relation is valid for any density matrix, and only relies on
the form of the time evolution operator and the weak field limit. 
In thermal equilibrium we can use the fluctuation-dissipation
relation to relate the retarded correlation function to the 
symmetric correlation function. In frequency space
\begin{align}
\label{FD}
  G^{AB}(\omega) = 
     \tanh\left(\frac{\omega}{2T}\right)
     G_S^{AB}(\omega) 
\end{align}
with $G^{AB}(t-t')\equiv \frac{1}{2}\langle [ \hat{A}(t),\hat{B}
(t')]\rangle$ and $G_S^{AB}(t-t')\equiv \frac{1}{2}\langle 
\{\hat{A}(t),\hat{B}(t')\}\rangle$. The Fourier transform of 
eq.~(\ref{Kubo_1}) then implies a Kubo relation in thermal
equilibrium, 
\begin{align}
\label{Kubo_FD} 
{\rm Im} \, \chi_{AB} (\omega) = 
\tanh\left(\frac{\omega}{2T}\right)\,  G_S^{AB}(\omega) . 
\end{align}
Kubo formulas for transport coefficients follow from the 
low frequency limit of the relation between the response
and the symmetric correlation function. In this limit we 
can approximate $\tanh(\omega/(2T))\simeq \omega/(2T)$ so 
that
\begin{align}
\label{Kubo_2}
 {\rm Im} \;\left.  
 \frac{\partial\langle\hat{A}(\omega)\rangle}{\partial h(\omega)}
     \right|_{h=0} = 
     \frac{\omega}{2T}\, G_S^{AB}(\omega) \, , 
      \hspace{0.5cm} (\omega\to 0). 
\end{align}
The most important example is the response of the stress tensor
to an external field that transforms like a metric perturbation. 
We consider $\hat{A}=\hat{B}=\hat{\Pi}^{ij}$ and 
\begin{align}
\hat{H}=\hat{H}_0 - \frac{1}{2}\int d^3x\, h_{ij}(t,x)
  \hat{\Pi}^{ij}(t,x)\, . 
\end{align}
The response function in frequency space for zero wave number 
$k=0$ is given by
\begin{align}
\label{resp_pi_xy}
   \chi^{xyxy}(\omega,0) = \left.
     \frac{\partial \langle \Pi^{xy}(\omega,0)\rangle}
      {\partial h_{xy}(\omega,0)}   \right|_{h_{xy}=0} \, . 
\end{align}
The usual assumption is that the low frequency limit of the 
response function can be computed using linearized deterministic 
fluid dynamics. In this case the response to a metric perturbation
$h_{xy}(\omega,0)$ can be determined using diffeomorphism 
invariance, in both the relativistic \cite{Baier:2007ix} and
the non-relativistic \cite{Son:2005tj} theory. The leading 
coupling to $h_{xy}$ arises from the derivative terms in the 
stress tensor, giving
\begin{align}
\label{pi_xy_resp_hydro}
 \Pi_{xy}(\omega,0)= i\omega\eta h_{xy}(\omega,0) + \ldots .
\end{align}
Combining eq.~(\ref{pi_xy_resp_hydro}) with eq.~(\ref{Kubo_2})
and eq.~(\ref{resp_pi_xy}) leads to the Kubo formula for the 
shear viscosity
\begin{align}
\label{Kubo_eta}
\eta = \frac{1}{2T}\int_{- \infty}^{\infty} dt \,  \int d^3x \, 
  \, \frac{1}{2} \langle \big\{ \hat{\Pi}_{xy}(0, \vec{0}), 
      \hat{\Pi}_{xy}(t, \vec{x})\big\} \rangle  \, . 
\end{align}
This result can be used to compute the viscosity starting 
from a microscopic quantum field theory, for example QCD. 
However, we can also make use of the fact that the RHS
is the zero frequency, zero wavenumber limit of the stress
tensor correlation function, which is a quantity that can 
be computed in a low-energy effective theory, such as 
kinetic theory or fluid dynamics. In this case we make 
use of the fact that the symmetric correlation function has
a straightforward classical limit, $\frac{1}{2}\langle\{ 
\hat{\Pi}_{xy}(0, \vec{0}),\hat{\Pi}_{xy}(t, \vec{x})\} 
\rangle \to \langle \Pi_{xy}(0, \vec{0})\Pi_{xy}
(t, \vec{x}) \rangle$. 

 In stochastic fluid dynamics the leading, tree-level, 
contribution to the statistical correlation function arises 
from the delta noise term in eq.~(\ref{noise-pi}). At this 
level the Kubo relation is a consistency check on the validity
of the fluctuation-dissipation relation. It confirms that the 
physical viscosity is equal to the bare viscosity. The fact 
that the integral in eq.~(\ref{Kubo_eta}) is dominated by 
a delta function reflects the fact that stochastic fluid 
dynamics is a low-energy effective theory. In this theory 
the microscopic origin of the bare viscosity is a UV phenomenon,
encoded in a contact term. 

 The full, renormalized viscosity can be computed by including 
the full stress tensor in eq.~(\ref{flux_piT}). In the lattice
discretized theory we calculate
\begin{align}
\label{Kubo_eta_lat}
    \eta = \frac{1}{T V} \, \int_0^{\infty} \, dt \, 
    \langle T_{xy}(0) \, T_{xy}(t) \rangle \, ,
\end{align}
where $T_{xy}(t)=\sum_x \Pi_{xy}(x,t)$ is the lattice sum of
the stress tensor, and we have made use of the symmetry of the 
correlation function under $t\to -t$. We note that the term 
$\nabla_{(i} \pi^T_{j)}$ vanishes when summed over the whole 
lattice. The remaining non-contact terms can be discretized as 
\begin{align}
\label{T_xy_lat}
  T_{xy}
    = \sum_{\vec{x}} 
    \left[
    \frac{1}{\rho} \pi_x^T(\vec{x})\pi_y^T(\vec{x}) +
    \, \frac{1}{4} \, 
    \left( \phi(\vec{x} + \hat{\mu}_x) - \phi(\vec{x} - \hat{\mu}_x) \right)
    \left( \phi(\vec{x} + \hat{\mu}_y) - \phi(\vec{x} - \hat{\mu}_y) \right)
    \right] \, ,
\end{align}
where $\hat{\mu}_i$ is a unit lattice vector in the $i$-direction.
A Kubo relation for the conductivity can be derived by studying the
response of the current to an external gauge field. Consider a
coupling of the form 
\begin{align}
    \hat{H}=\hat{H}_0 - \int d^3x\, A^i(x,t)
       \hat{\jmath}_i(x,t)\, . 
\end{align}
The linear response to $\vec{A}$ leads to the Kubo relation for the 
conductivity. In the classical limit we have  
\begin{align}
\label{Kubo_kappa}
    \kappa = \frac{1}{3TV} \, \int_0^{\infty} \, dt \, 
    \langle J_{i}(0) \, J^i(t) \rangle \, .
\end{align}
Here, $J^i(t)=\sum_x \jmath^i(x,t)$ is the lattice sum of the current.
When summed over the entire lattice the gradient term $\jmath_i \sim 
\nabla_i [(\delta {\cal H})/(\delta \phi)]$ does not contribute to
$J_i$. The discretized current is given by 
\begin{align}
  J_i = \frac{1}{\rho}\,  \sum_x  \, \phi(\vec{x})\pi_i^T(\vec{x}). 
\end{align}

\begin{figure}[t!]
\subfloat[]{
\includegraphics[width=0.30\columnwidth]{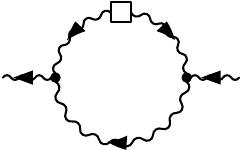}}
\hspace{0.1\columnwidth}
\subfloat[]{
\includegraphics[width=0.30\columnwidth]{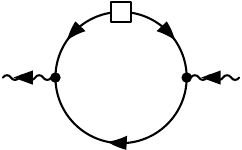}}
\\
\subfloat[]{
\includegraphics[width=0.3\columnwidth]{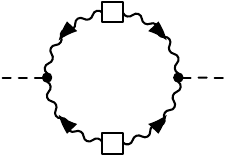}}
\hspace{0.1\columnwidth}
\subfloat[]{
\includegraphics[width=0.3\columnwidth]{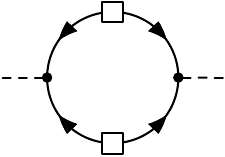}}
\caption{Diagrammatic representation of one-loop corrections 
to the shear viscosity. Fig.~(a) and (b) show the self-advection
and order parameter contributions to the retarded self energy 
of the momentum density. Fig.~(c) and (d) show the corresponding
contributions to the Kubo integrand. The dashed line represents
an insertion of the stress tensor.
\label{fig:1loop-visc}}
\end{figure}

\section{Theoretical Expectations}
\label{sec:theory}
\subsection{Non-critical fluids}
\label{sec:theory-pt}

 In the non-critical regime the response of the fluid to small 
perturbation can be computed order by order in a low frequency, 
long wavelength expansion. In thermal equilibrium the retarded 
correlation function of the momentum density $\pi_i^T$ is
given by 
\begin{align}
 G_R^{ij}(\omega,k) = \frac{P_T^{ij}(k)}
      {-i\omega+D^0_\eta k^2+\Sigma(\omega,k)}\, , 
\end{align}
where $D^0_\eta=\eta_0/\rho$ is the bare momentum diffusion constant,
$P^{ij}_T=\delta^{ij}-\hat{k}^i\hat{k}^j$ is the transverse projection
operator, and $\Sigma(\omega,k)$ is the self energy. The correction
to the shear viscosity is 
\begin{align}
    \delta\eta = \rho \, 
      \left.\frac{\partial \Sigma(0,k^2)}{\partial k^2}\right|_{k^2=0}.
\end{align}
The leading one-loop diagrams that contribute to the self energy of the
momentum density in model H are shown in Fig.~\ref{fig:1loop-visc}(a)
and (b). Here, wavy lines denote Green functions of $\pi_i^T$, solid
lines are Green functions of $\phi$, an open box is a noise insertion,
and the vertices correspond to the self advection of $\pi_i^T$ and the
advection of $\phi$ by the momentum density. The arrows indicate the
causal structure. A Green function with one arrow is retarded, and 
a Green function with two out-going arrows is symmetric. The coupling
between $\phi$ and $\pi_i^T$ contains additional powers of the momentum
compared to the self-coupling of $\pi_i^T$ and the diagram in
Fig~\ref{fig:1loop-visc}(b) is suppressed with respect to the diagram 
in (a). The correction to the shear viscosity is UV divergent. This
divergence is typically regularized by performing the frequency integral
first, and then cutting off the integral over wave number at a scale 
$\Lambda_{\it UV}$. The result in three dimensions is
\cite{Kovtun:2012rj,Chao:2020kcf}
\begin{align} 
\label{del-eta-sig}
\delta \eta = \frac{7}{60\pi^2}\frac{\rho T\Lambda_{\it UV}}{\eta_0}\, . 
\end{align}
At non-zero frequency the self-energy contains non-analytic 
terms, $\partial_{k^2}\Sigma(\omega,k^2)|_{k^2=0}\sim \sqrt{\omega}$.
In two spatial dimensions the correction to the viscosity is 
logarithmically divergent \cite{Kadanoff:1989,Kovtun:2012rj},
\begin{align} 
\label{del-eta-2d}
\delta \eta(\omega) = \frac{1}{32\pi}\frac{\rho T}{\eta_0}
\log\left( \frac{\sqrt{2}\Lambda_{\it UV}^2\eta_0}
                                 {\omega\rho}\right)\, ,
\end{align}
and the limit $\omega\to 0$ does not exist. 
This result implies that, strictly speaking, the Navier-Stokes
equation is not valid in two dimensional fluids, and one has to work 
with stochastic fluid dynamics, even at leading order and away 
from the critical point. In practice, the logarithm is 
rendered finite because of the finite system size and 
observation time, and an approximate Navier-Stokes theory 
may be useful.

\begin{figure}[t!]
\subfloat[]{
\includegraphics[width=0.30\columnwidth]{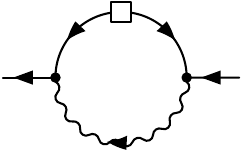}}
\hspace{0.01\columnwidth}
\subfloat[]{
\includegraphics[width=0.30\columnwidth]{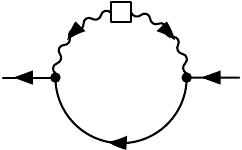}}
\hspace{0.01\columnwidth}
\subfloat[]{
\includegraphics[width=0.27\columnwidth]{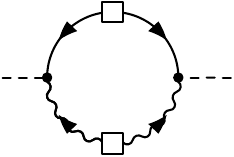}}
\caption{Diagrammatic representation of one-loop corrections to the
conductivity. Fig.~(a) and (b) show the advection contributions to 
the retarded self energy of the order parameter. Fig.~(c) shows the
corresponding contributions to the Kubo integrand. The dashed line 
represents an insertion of the current.
\label{fig:1loop-cond}}
\end{figure}

 The retarded Green function of the order parameter field is 
\begin{align}
 S_R(\omega,k) = \frac{1}
      {-i\omega+D^0_\kappa k^2+\Xi(\omega,k)}\, , 
\end{align}
where $D_\kappa=\kappa\chi_\phi^{-1}$ and $\Xi(\omega,k)$ is the 
self energy. Fig.~\ref{fig:1loop-cond}(a) and (b) show the one-loop 
contributions to the self energy. We again find that there is a 
linear divergence in three dimensions, and a logarithmic term
in two dimensions. The correction to the diffusion constant in 
$d=3$ is \cite{Kovtun:2003vj,Kovtun:2012rj,Martinez:2018wia}
\begin{align}
\label{del-D}
   \delta D_\kappa = \frac{1}{3\pi^2}
      \frac{T\Lambda_{\it UV}}{\rho(D_\eta^0+D_\kappa^0)} \, , 
\end{align}
and in two dimensions 
\begin{align}
\label{del-D-2d}
   \delta D_\kappa(\omega) = \frac{1}{8\pi}
      \frac{T\Lambda_{\it UV}}{\rho(D_\eta^0+D_\kappa^0)} \,
      \log\left(\frac{\Lambda_{\it UV}^2(D^0_\eta+D^0_\kappa)}
               {\sqrt{2}\,\omega}\right)
      \, .
\end{align}
These calculations can be compared to results based on 
the Kubo relations, eq.~(\ref{Kubo_eta},\ref{Kubo_kappa}).
The one-loop diagrams that contribute to the shear viscosity
are shown in Fig.~\ref{fig:1loop-visc}(c) and (d). As before 
the first diagram dominates, and contains a UV divergence.
This divergence can be regularized in the same way as the 
one-loop self energy. The Kubo integral at finite frequency 
is given by \cite{Kovtun:2011np,Chafin:2012eq}
\begin{align}
\label{del-eta-Kubo}
\eta(\omega) = \eta_0
   + \frac{7}{60\pi^2}\frac{\rho T\Lambda_{\it UV}}{\eta_0}
   - \frac{7}{240\pi}\frac{\rho^{3/2} T}{\eta_0^{3/2}}\, \sqrt{\omega}
   + \ldots \, , 
\end{align}
where the ellipsis refer to terms of higher order in frequency. 
We observe that the calculation based on the Kubo relation
agrees with that based on the self energy. This is a non-trivial
check: The one-loop diagrams in \ref{fig:1loop-visc}(a) and 
(c) involve different vertices, and different Green functions
\footnote{Note that many calculations of the Kubo integral use 
the fluctuation-dissipation theorem to relate the diagram 
in Fig.~\ref{fig:1loop-visc}(c) to a retarded function, see
for example \cite{Kovtun:2011np}.}.
The one-loop Kubo integral for the conductivity, shown in
Fig.~\ref{fig:1loop-cond}(c), has not been computed in the 
literature, but the agreement with eqs.~(\ref{del-D}) and 
(\ref{del-D-2d}) can be checked using the methods described 
in \cite{Kovtun:2011np,Chafin:2012eq}.

 Finally, we can use perturbative calculations to determine how 
the infrared divergence in two dimensions is regularized in a 
finite volume. In a finite volume there are IR and UV cutoffs 
on the momenta, $\Lambda_{\it IR}=\pi/(La)$ and $\Lambda_{\it UV}
=\pi/a$. The IR sensitive contribution to the shear viscosity is 
\begin{align} 
\label{del-eta-2d-L}
\delta \eta(L) = \frac{1}{16\pi}\frac{\rho T}{\eta_0}
\log\left( L \right)\, ,
\end{align}
and there is an analogous $\log(L)$ dependence in the diffusion 
constant.

\subsection{Critical fluids}
\label{sec:theory-crit}

 Critical behavior is most directly reflected in the order parameter
correlation function. Near the critical point the correlator is 
non-perturbative, but there is a simple model, known as the ``mode
coupling'' or Kawasaki approximation  \cite{Kawasaki:1970}, that 
describes the behavior of the correlation function quite well. 
In the Kawasaki approximation the self energy is independent of 
frequency $\Xi(\omega,k)\simeq \Gamma_k$ \footnote{
Note that at any order in perturbation theory the
self energy satisfies $\Xi(\omega,q=0)=0$, see, for example, 
reference \cite{Tauber:2014}.}. 
Here, the relaxation rate $\Gamma_k$ is given by 
\begin{align}
\Gamma_k = \frac{\kappa}{\xi^4} 
      \left(k\xi\right)^2 \left(1+(k\xi)^2\right)
     + \frac{T}{6\pi\eta_R\xi^3}\, K(k\xi)\, , 
\label{Gamk:Kaw}
\end{align}
and the Kawasaki function is \cite{Kawasaki:1970,Onuki:2002}
\begin{align}
   K(x) = \frac{3}{4} \left[ 
     1+x^2+\left(x^3-x^{-1}\right)\tan^{-1}(x)\right]\, .
\label{Kaw}
\end{align}
This function behaves as $K(x)\simeq x^2$ for $x\ll 1$ and $K(x)
\simeq (3\pi/8)x^3$ for $x\gg 1$. The first term in eq.~(\ref{Gamk:Kaw}) 
corresponds to model B behavior, $\Gamma_k\simeq \kappa\, \chi_\phi^{-1} 
k^2$, where the susceptibility has the Ornstein-Zernike form $\chi_\phi = 
\xi^2/(1+(k\chi)^2$. The second term encodes the critical behavior in 
model H. If the correlation length is large, $\xi\gg (6\pi \eta_R\kappa)
/T$, then the small momentum behavior is $\Gamma_k \simeq T k^2/(6\pi
\eta_R\xi)$. If we define the renormalized conductivity $\kappa_R$ by
$\Gamma_k=\kappa_R\chi_\phi^{-1}k^2$ then we conclude that $\kappa_R
\sim \xi^{x_\kappa}$ with $x_\kappa=1$.

 Within this model we have assumed that the renormalized viscosity 
$\eta_R$ is approximately constant. This assumption can be checked 
by computing the diagram in Fig.~\ref{fig:1loop-visc}(b) using 
the full propagator with the self energy $\Xi(\omega,k)\simeq 
\Gamma_k$ included. This calculation gives a weak divergence of
the viscosity in $d=3$ dimensions, 
\begin{align}
\label{eta-crit}
\eta_R \sim \xi^{x_\eta}\, , 
\hspace{1cm}
x_\eta=\frac{8}{15\pi^2}\simeq 0.054\, . 
\end{align}
implying that the Kawasaki approximation is approximately 
self-consistent. The critical exponents $x_\kappa$ and $x_\eta$
satisfy a scaling relation 
\begin{align}
    x_\kappa+x_\eta = 4-d-\eta^*\, , 
\label{x-kappa-eta}    
\end{align}
where $\eta^*$ is the static correlation function exponent. 
Historically, this relation was first obtained from a physical
argument described in \cite{Arcovito:1969,Hohenberg:1977ym}.
Using renormalization group arguments it was also shown 
that eq.~(\ref{x-kappa-eta}) follows from the symmetries 
of the stochastic effective action \cite{Siggia:1976}. 
These symmetries also imply a relation between the dynamical
critical exponent $z$ and the scaling of the transport 
coefficients
\begin{align}
   z=4-x_\kappa-\eta^*\, .
\label{z-Gam}   
\end{align}
The Kawasaki approximation can also be used to determine 
the frequency dependence of the viscosity. The result is 
\cite{Ohta:1976}
\begin{align}
\label{eta_om_crit}
\eta(\omega) \sim (i\omega t_\xi)^{-x_\eta/3}\, , 
\hspace{0.5cm}
 \omega t_\xi\gg 1\, ,
\end{align}
where $t_\xi \simeq 6\pi\eta_R\xi^3/T$. The frequency 
dependence is even weaker than the dependence on the 
correlation length. We also note that eq.~(\ref{eta_om_crit})
implies that the asymptotics of the Kubo integrand is only 
reached at very large times, $t>t_\xi\sim \xi^3$.

More accurate values of the $x_\kappa$ and $x_\eta$ can be 
obtained from the $\epsilon$ expansion. At two loops we have
\cite{Adzhemyan:1999h}
\begin{align}
\label{crit-eps}
\begin{array}{c|cccc}
      & x_\kappa  &   x_\eta     &  \eta^*          &   z   \\ \hline 
  \;d=2\; & \;\; 1.57\;\; &  \;\;0.18\;\;&  \;\;0.250\; \; & \;\;2.18\;\;  \\
  d=3 & 0.89      &   0.07       &  0.0363          &  3.07     
\end{array}
\end{align}
where we have used the more accurate value of $\eta^*$ from the 
numerical bootstrap \cite{Alday:2015ota} (in $d=2$ the value of 
$\eta^*$ is exact). Critical exponents have also been determined 
using the functional renormalization group. In $d=3$ the results 
are close to those of the epsilon expansion, but in $d=2$ the 
functional RG predicts either zero \cite{Roth:2024hcu} or very 
small values of $x_\eta$ \cite{Chen:2024lzz}.

\section{Numerical results}
\label{sec:num}

\begin{figure}[t]
\begin{center}
\includegraphics[width=0.45\columnwidth]{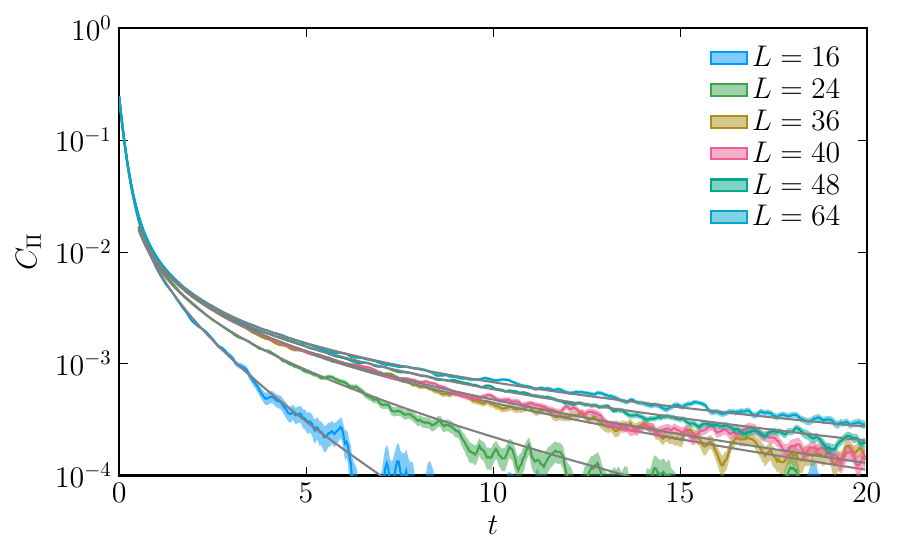}
\includegraphics[width=0.45\columnwidth]{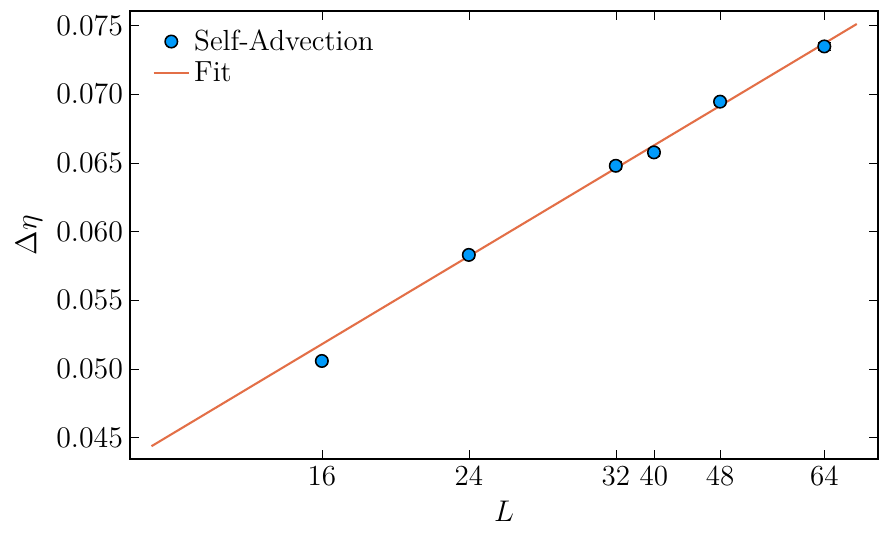}
\end{center}
\caption{Left panel: Correlation function $C_\Pi(t)$ of the stress 
tensor $T_{xy} =  \pi_x \pi_y/\rho$ in a 2d non-critical fluid. 
The dynamics only contains the self-advection term, and the bare
shear viscosity is $\eta_0=1$. The correlation function is shown
in several different volumes $V=(La)^2$ with $L=16,24,36,40,48,64$.
The gray curves show numerical fits $C_\Pi(t)\sim A \exp(- pt)/t$.   
Right panel: Scaling of the Kubo integral for viscosity with system
size $L$. The orange line shows a fit of the form $\Delta\eta= \alpha
\log(L)$ with $\alpha=0.0158$.
\label{fig:eta-2d-non-crit}}
\end{figure}

\subsection{Two-dimensional fluids: Non-critical case}
\label{sec:num-2d}

 We first study how shear viscosity is modified by fluctuations 
in a non-critical two-dimensional fluid. Our goal is to verify the 
logarithmic divergence of $\eta$ with system size predicted in 
eq.~(\ref{del-eta-2d-L}). We compute $\delta\eta$ using the Kubo
relation in eq.~(\ref{Kubo_eta_lat}). We focus on the contribution 
from self-advection. In this case $T_{xy}$ only contains the first 
term in eq.~(\ref{T_xy_lat}), and the only non-linearity in the 
equations of motion is the self-coupling of $\pi_i^T$. We do not 
include the noise term in $T_{xy}$. The Metropolis algorithm ensures
that the contact term in the Kubo relation exactly reproduces the bare 
viscosity. 

 The Kubo integrand $C_\Pi(t)=\langle T_{xy}(0) T_{xy}(t)\rangle$ 
computed in different volumes $V=(La)^3$ is shown in the left panel of 
Fig.~\ref{fig:eta-2d-non-crit}.  Here, we have chosen $\eta_0=1$ and 
$\rho=1$ in lattice units, see Sect.~III.E in \cite{Chattopadhyay:2024bcv}.
The result is consistent with the expectation that in the infinite volume 
and continuum limits $C_\Pi(t)\sim 1/t$, but that in a finite system the 
UV and IR divergences are cut off at a scale $t_\Lambda \sim (D_\eta 
\Lambda^2)^{-1}$ with $\Lambda_{\it UV}\sim a^{-1}$ and $\Lambda_{\it IR}
\sim (La)^{-1}$. The Kubo integral $\delta\eta= (VT)^{-1}\int dt\, 
C_\Pi(t)$ is shown in the right panel. We clearly observe the scaling 
with $\log(L)$. The best-fit coefficient of the logarithm is 0.0158, 
which should be compared with the prediction in eq.~(\ref{del-eta-2d-L}), 
$1/(16\pi)\simeq 0.0198$ (for $\rho=\eta_0=1$). Given that this result 
is based on computing the diagram in the continuum, not using lattice 
Green functions, the agreement is quite good. 

\subsection{Non-critical fluids in three dimensions}
\label{sec:num-3d}

\begin{figure}[t]
\begin{center}
\includegraphics[width=0.6\columnwidth]{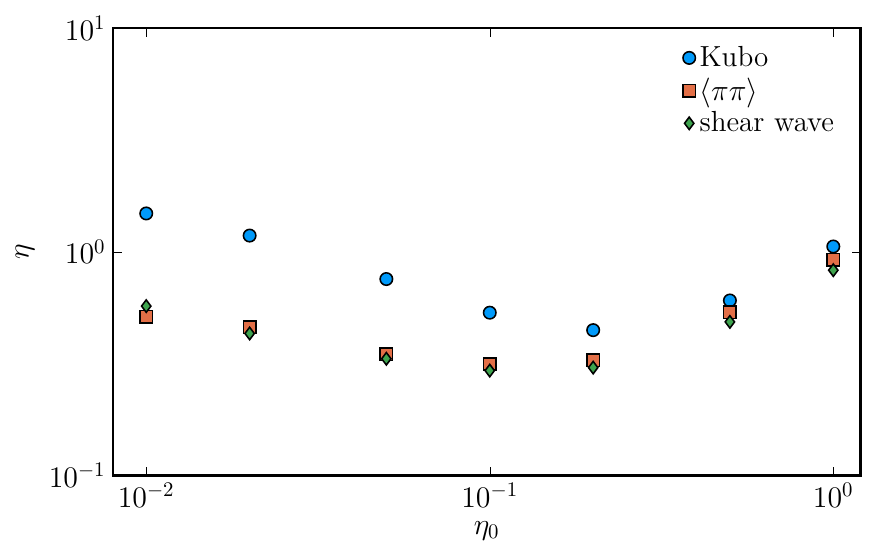}
\end{center}
\caption{Shear viscosity of a non-critical model H fluid
in three dimensions, measured in a volume $V=(La)^3$ with 
$L=24$. We show the physical shear viscosity as a function 
of the bare value. The viscosity is extracted using the 
momentum density correlator (red squares), the decay of a shear 
wave (green diamonds), and the Kubo formula (blue circles). 
\label{fig:eta-plat-Kubo}}
\end{figure}

  In three dimensions it is possible to extract the shear viscosity 
from the exponential decay of the correlation function of the momentum 
density, see Sect.~\ref{sec:visc} of this work and Fig.~6 in
\cite{Chattopadhyay:2024bcv}. As a consistency check we have 
also determined the shear viscosity using the decay of a shear
wave. For this purpose we select an initial condition that corresponds
to a shear wave $\pi_x(t=0)=\pi_x^0 \sin(ky)$ with wave number $k$.
We average the profile over many trajectories with the same initial
condition, and extract the viscosity from the exponential decay of 
the amplitude \footnote{Another numerical viscometer that has been
considered in the literature is based on the mean velocity profile 
observed in sheared (Poiseuille) flow \cite{Kadanoff:1989}}. 
In Fig.~\ref{fig:eta-plat-Kubo} we compare the results to those 
obtained from the correlation function of $\vec\pi^T$. We find that
there is good agreement between the two methods. For large values 
of $\eta_0$ the physical viscosity is close to the bare one. As 
$\eta_0$ is reduced the fluctuation induced contribution grows,
consistent with the perturbative expectation $\delta\eta \sim 
1/\eta_0$, see eq.~(\ref{del-eta-sig}). The physical shear 
viscosity has a minimum as a function of $\eta_0$, and remains 
finite in the limit $\eta_0\to 0$. 

 Note that all quantities are measured in lattice units, so 
that we cannot directly observe the linear divergence of $\delta
\eta$ with $a^{-1}$. On the lattice, time is measured in units 
of $t_{\it lat} = a^4/\kappa$. This implies that shear viscosity 
is given in units of $\eta_{\it lat}=T_0a/\kappa$, where $T_0$ 
is the unit of temperature, and that mass density or enthalpy 
density are measured in units of $\rho_{\it lat}=T_0a^3/\kappa^2$. 
The fact that $\delta\eta\sim a^{-1}$ for fixed physical values 
of $\eta_0$ and $\rho$ is implied by the observed scaling 
$\delta\eta \sim \rho/\eta_0$. This follows from
\begin{align}
  \delta\eta \;\sim\; 
  \frac{(\eta_0/\eta_{\it lat})}{(\rho/\rho_{\it lat})}
  \, (T/T_0) \, \eta_{\it lat}
  \;\sim\; \frac{\rho T}{\eta_0} 
    \, a^{-1}\, . 
\end{align}
 In Fig.~\ref{fig:eta-plat-Kubo} we compare measurements of 
the viscosity using the correlation function $C_\pi(t,k)$ with 
results based on the Kubo formula. We have included the contact
term due to the noise in the Kubo integrand so that the results 
for large $\eta_0$ match. We observed that while the functional 
dependence of the viscosity $\eta$ on the bare value $\eta_0$ is 
very similar, the overall magnitude of $\delta\eta$ differs by 
a factor $\sim 2$. We showed in eq.~(\ref{del-eta-Kubo}) that,
if the self energy and the Kubo integrand in the continuum 
limit are regularized in  a consistent way, then the two
calculations of $\delta\eta$ agree. This suggests that there is 
an inconsistency in the UV behavior of the lattice regularized 
correlation functions. As discussed in Sect.~\ref{sec:num-2d} the 
lattice regulates wave numbers at a scale $\Lambda_{\it UV}\sim
a^{-1}$. In the Kubo integrand this corresponds to a short-time 
cutoff $t_{\it UV}\sim a^2/D_\eta$.

 The stochastic algorithm involves two additional UV scales. 
One is given by the time step in the advection/diffusion 
algorithm. A typical time step is $\Delta t=0.04\, t_{\it lat}$.
This should be compared to the UV scale defined in the previous
paragraph, which can be written as $t_{\it UV} = (\eta_{\it 
lat}/\eta)\, t_{\it lat}$, where we have assumed that $\rho/
\rho_{\it lat}=1$. This implies that the time step is not a 
concern, unless $\eta$ is very large. Indeed, we have checked 
that the discrepancy in Fig.~\ref{fig:eta-plat-Kubo} is 
unaffected by the value of $\Delta t$.

\begin{figure}[t]
\begin{center}
\includegraphics[width=0.48\columnwidth]{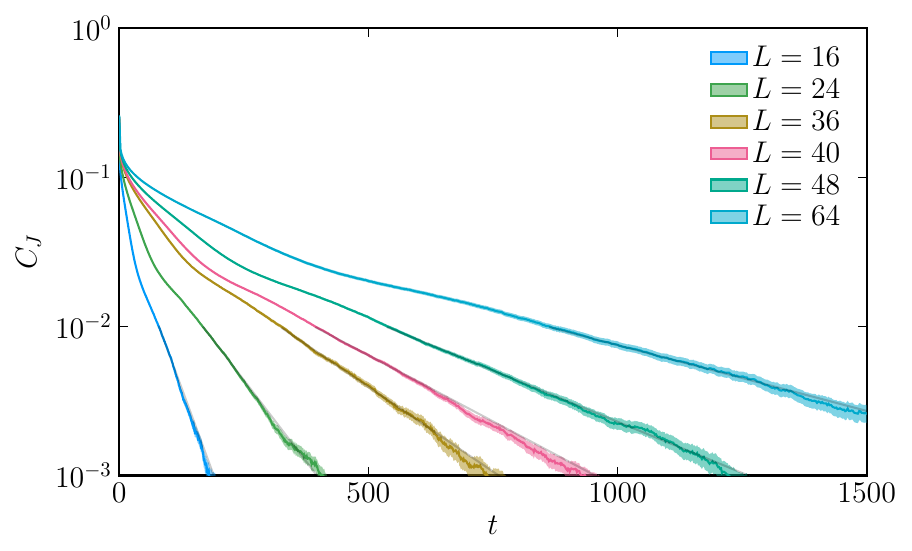}
\includegraphics[width=0.45\columnwidth]{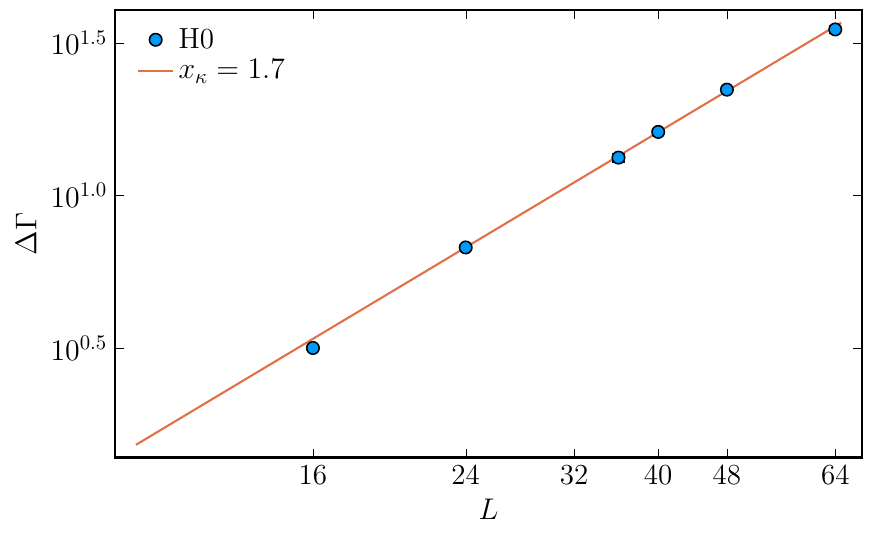}
\end{center}
\caption{
\label{fig:kappa-2d-crit}
Thermal conductivity in a critical two-dimensional fluid 
governed by model H0. Left panel: Kubo integrand $C_\jmath(t)$
in different volumes $V=(La)^2$ for $L=16,24,\ldots,64$. Right
panel: Kubo integral $\Delta\kappa$ as a function of system
size, together with the best fit $\Delta\kappa\sim L^{x_\kappa}$
where $x_\kappa=1.7$.}
\begin{center}
\includegraphics[width=0.45\columnwidth]{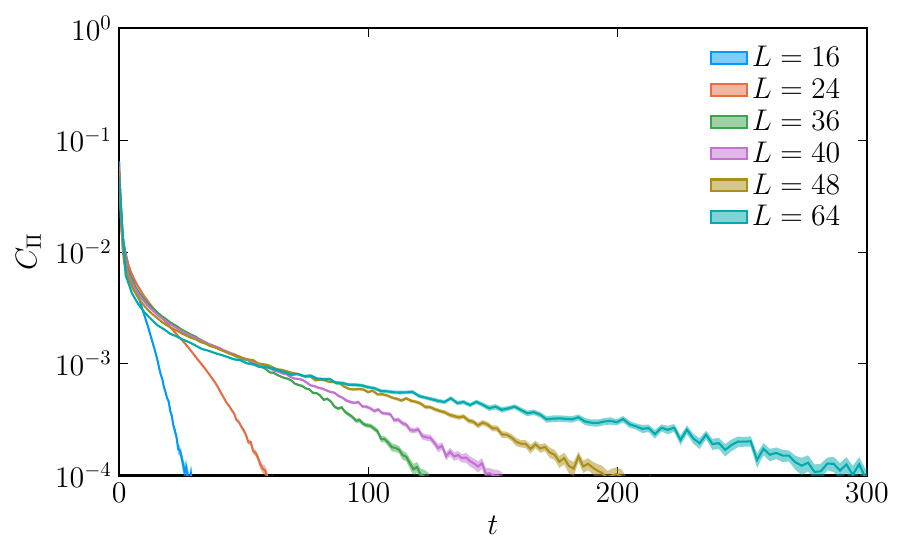}
\includegraphics[width=0.45\columnwidth]{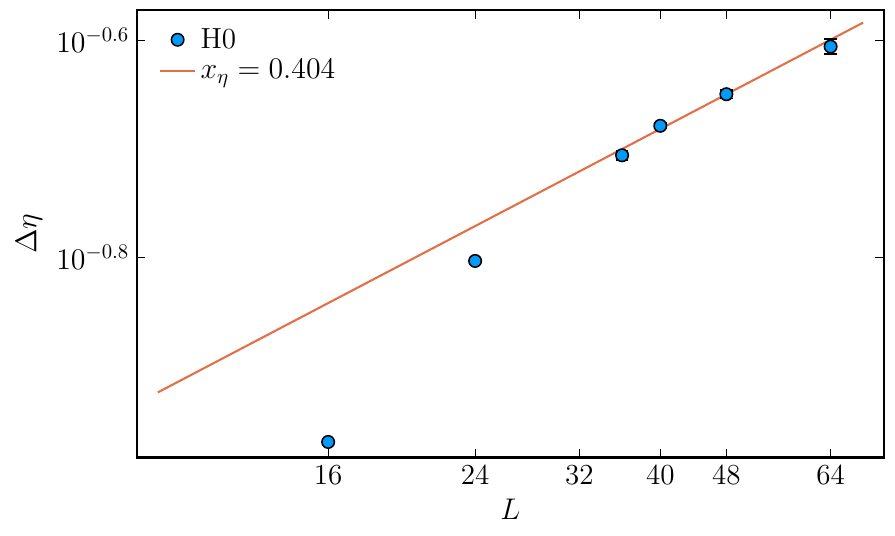}
\end{center}
\caption{\label{fig:eta-2d-crit}
Shear viscosity in a critical two-dimensional fluid governed 
by model H0. Left panel: Kubo integrand $C_\Pi(t)$ in different 
volumes $V=(La)^2$ for $L=16,24,\ldots,64$. Right panel: Kubo 
integral $\Delta\eta$ as a function of system size, together 
with the best fit $\Delta\eta\sim L^{x_\eta}$ where $x_\eta
=0.404$.}
\end{figure}

 The second UV scale is associated with the advection term in 
the equation of motion. In the continuum limit, the advection
term is of the form $\partial_t \pi^T_i\sim \frac{1}{\rho}
(\pi^T_j k_j^{\mbox{}})\pi^T_i$. On the lattice, this vertex
is modified in a way that depends on the scheme used to discretize
the advection term. For a centered derivative, $k_i\to\tilde{k}_i$
with $\tilde{k}_i=a^{-1}\sin(k_ia)$. This means that for momenta
at the cutoff scale, $k\sim \Lambda_{\it UV}$, the advection term
is strongly modified. We have checked that for a non-critical 
fluid in the regime where $\delta\eta\lsim \eta_0$ the Kubo 
integral is not very sensitive to the advection term. Indeed, 
the integral is well approximated by the result in a theory
without mode couplings. However, the advection term controls the
renormalization of $\eta$ extracted from the correlation function
of $\pi^T_i$. In particular, $\delta\eta$ vanishes in a theory
without mode couplings \footnote{
One may ask whether it is possible to check the effect of 
different discretization of the advection term. This is not 
straightforward to do, since the specific discretization used 
in this work, eq.~(42) in \cite{Chattopadhyay:2024bcv}, is 
distinguished by the fact that the discretized equations
conserve energy.}. 
  
We conclude that the Kubo formula, at least in the form stated in
eq.~(\ref{Kubo_eta_lat},\ref{T_xy_lat}), cannot be used to extract 
the UV sensitive part of $\delta\eta$. This means that the Kubo 
formula cannot be used to determine the relation between the bare
viscosity $\eta$ and the renormalized viscosity $\eta_R$. However, 
Fig.~\ref{fig:eta-2d-non-crit} shows that the Kubo formula correctly 
reproduces the IR contribution to the viscosity. As a result, we 
expect that the Kubo relation can be used to extract the critical 
contribution to transport coefficients.

\subsection{Critical transport in two dimensions}
\label{sec:num-2d-crit}

  We first study transport coefficients in a two-dimensional 
critical fluid. We will consider two theories, model H as well
as a truncation that we refer to as model H0. In model H0 the 
self-advection term of the momentum density is dropped. As 
we have seen, this implies that the UV sensitive additive 
renormalization of the shear viscosity is significantly smaller. 
We have previously demonstrated, however, that this shift does 
not affect universal predictions of critical fluid dynamics
\cite{Chattopadhyay:2024bcv}. In particular, the dynamical 
critical exponent $z$ is the same in model H and model H0.
Indeed, we observed an even stronger version of 
universality: For a fixed correlation length, and for the same 
value of the physical viscosity, the scaling exponent in models 
H and H0 is identical, see Fig.~9 in \cite{Chattopadhyay:2024bcv}.
We also observed, however, that genuine model H scaling 
requires $\xi\gsim (6\pi\kappa\eta_R)/T$, and that this regime 
is easier to reach in model H0, because the renormalized
value of the viscosity can be made smaller.

 Fig.~\ref{fig:kappa-2d-crit} shows the critical scaling of 
the thermal conductivity in model H0. The left panel presents 
the Kubo integrand $C_\jmath(t)$ at the critical point in 
different volumes $V=(La)^3$. The right panel shows the Kubo
integral $\Delta\kappa=(3VT)^{-1}\int dt\, C_{\jmath}(t)$ as 
a function of $L$ in a double logarithmic plot. We also show
the best fit $\Delta\kappa\sim L^{x_\kappa}$ with $x_\kappa=
1.70$. We observe that the data clearly exhibit critical 
scaling, that scaling sets already in the smallest volume 
studied, and that the critical divergence is indeed associated
with the long-time behavior of $C_\jmath(t)$. 

  Fig.~\ref{fig:eta-2d-crit} shows the scaling of the shear 
viscosity, also in model H0. Note that in this theory the 
stress tensor only contains the contribution from the order 
parameter, $\Pi_{ij} = \nabla_i\phi\nabla_j\phi$. Again 
we observe critical scaling, although scaling behavior only
sets in at somewhat larger values of $L$. The best fit value 
of the critical exponent is $x_\eta=0.404$. If this number
is combined with the value of $x_\kappa$ we obtain $x_\eta
+x_\kappa\approx 2.1$ which should be compared to right hand side
of the scaling relation in eq.~(\ref{x-kappa-eta}), $2-\eta^*
=1.75$. Note that we have not attempted to compute the value 
of $\eta^*$, but have used the exact value. 

\begin{figure}[t]
\begin{center}
\includegraphics[width=0.45\columnwidth]{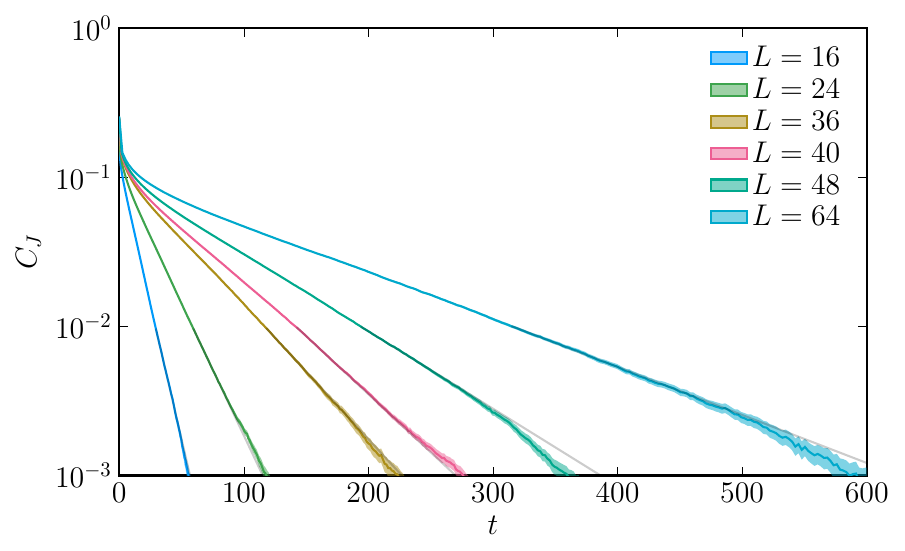}
\includegraphics[width=0.45\columnwidth]{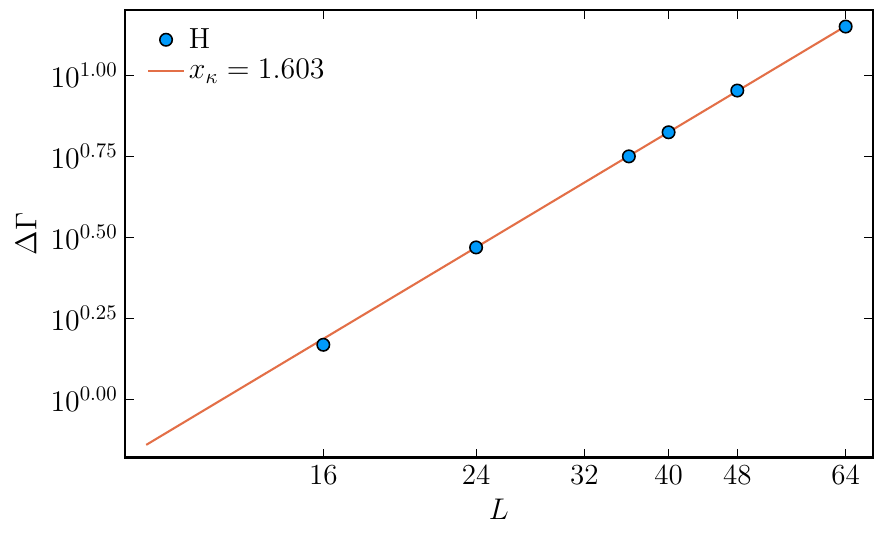}
\end{center}
\caption{\label{fig:kappa-2d-crit-H}
Same as Fig.~\ref{fig:kappa-2d-crit} in model H. Left panel
shows the Kubo integrand $C_\jmath(t)$ as a function of $t$, 
and the right panel presents $\Delta\kappa$ as a function
of $L$. Also shown is the best fit $\Delta\kappa\sim L^{x_\kappa}$ 
with $x_\kappa=1.603$.}
\begin{center}
\includegraphics[width=0.45\columnwidth]{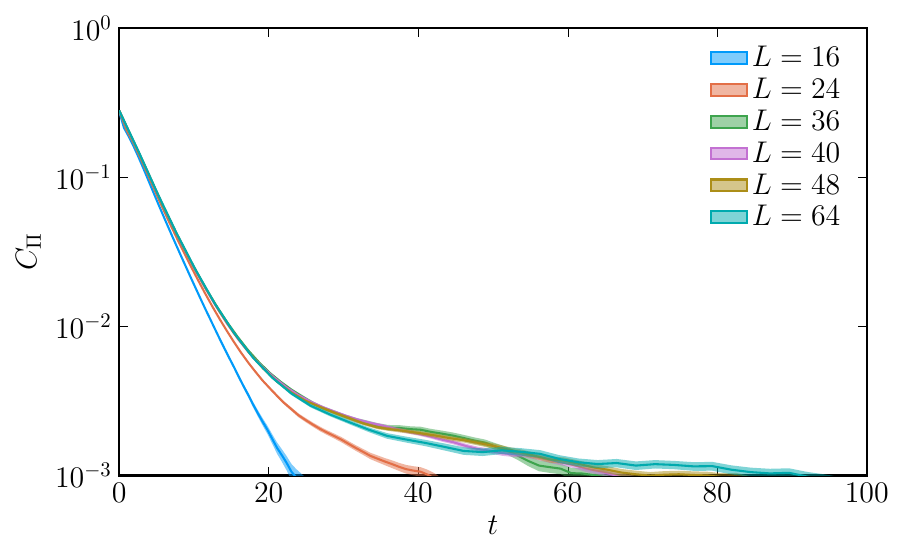}
\includegraphics[width=0.45\columnwidth]{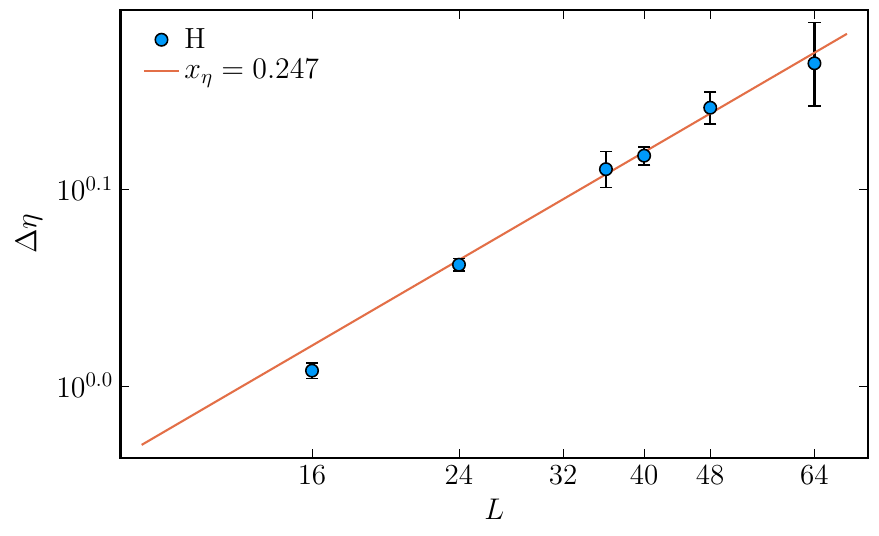}
\end{center}
\caption{\label{fig:eta-2d-crit-H}
Same as Fig.~\ref{fig:eta-2d-crit} in model H. Left panel
presents the Kubo integrand $C_\Pi(t)$ as a function of $t$, 
and the right panel displays $\Delta\eta$ as a function
of $L$. Also shown is the best fit $\Delta\eta\sim L^{x_\eta}$ 
with $x_\eta=0.247$. } 
\end{figure}

  We observe that the Monte Carlo errors on individual data 
points in Figs.~\ref{fig:kappa-2d-crit} and \ref{fig:eta-2d-crit}
are small, but systematic errors are harder to assess. Clearly, 
there is an uncertainty related to the fit window in $\log(L)$,
which is more serious in the case of $x_\eta$. There is also an 
uncertainty due to incomplete equilibration, in particular when 
we have to determine the integral over a long-time tail, as is 
the case here. As a conservative approach we will assume that 
the discrepancy in the scaling relation is related to systematic
errors. In that case, errors in $x_\kappa$ and $x_\eta$ are of 
order $\Delta x_{\kappa,\eta}\simeq 0.2$.

 Fig.~\ref{fig:kappa-2d-crit-H} and \ref{fig:eta-2d-crit-H}
show the corresponding results in model H. Again, critical 
scaling is clearly observed, although the error bars are 
somewhat larger as compared to model H0. We find $x_\kappa=1.603$
and $x_\eta=0.247$. This implies $x_\kappa+x_\eta=1.85$, in
better agreement with the prediction of the scaling relation, 1.75.
In model H the second scaling relation, eq.~(\ref{z-Gam}), is 
given by 
\begin{align}
z = 2.11 \hspace{1cm}
4-x_\kappa - \eta^* = 2.15\, ,
\end{align}
where we have used the value of $z$ determined in 
\cite{Chattopadhyay:2024bcv}. We observe that in model H the 
scaling window for $x_\eta$ is somewhat larger as compared to
model H0, consistent with the fact that the deviation in the 
scaling relations is smaller. We conclude that systematic 
errors in $x_\kappa$ and $x_\eta$ are also reduced, on the 
order of $\Delta x_{\kappa,\eta}\simeq 0.1$. This implies
that, within our systematic errors, the critical exponents
$x_\kappa$ and $x_\eta$ are the same in model H0 and model H.

\begin{figure}[t]
\begin{center}
\includegraphics[width=0.45\columnwidth]{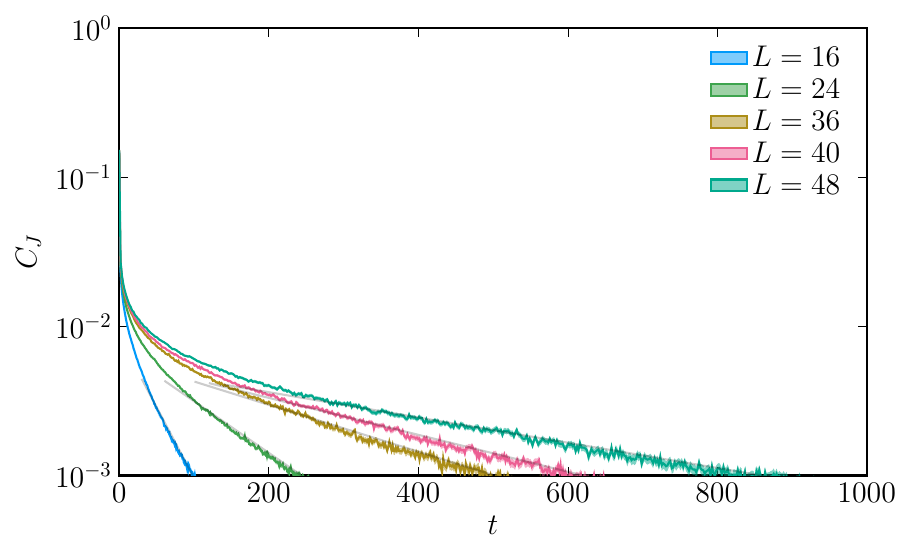}
\includegraphics[width=0.45\columnwidth]{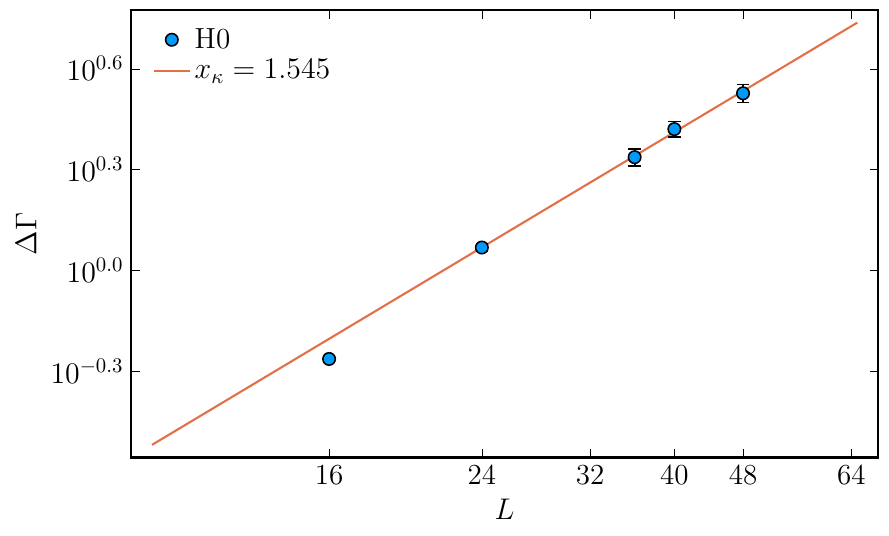}
\end{center}
\caption{\label{fig:kappa-3d-crit}
Thermal conductivity in a critical three-dimensional fluid 
governed by model H0. Left panel: Kubo integrand $C_\jmath(t)$
in different volumes $V=(La)^3$ for $L=16,24,\ldots,48$. Right
panel: Kubo integral $\Delta\kappa$ as a function of system
size, together with the best fit $\Delta\kappa\sim L^{x_\kappa}$
where $x_\kappa=1.545$.}
\begin{center}
\includegraphics[width=0.45\columnwidth]{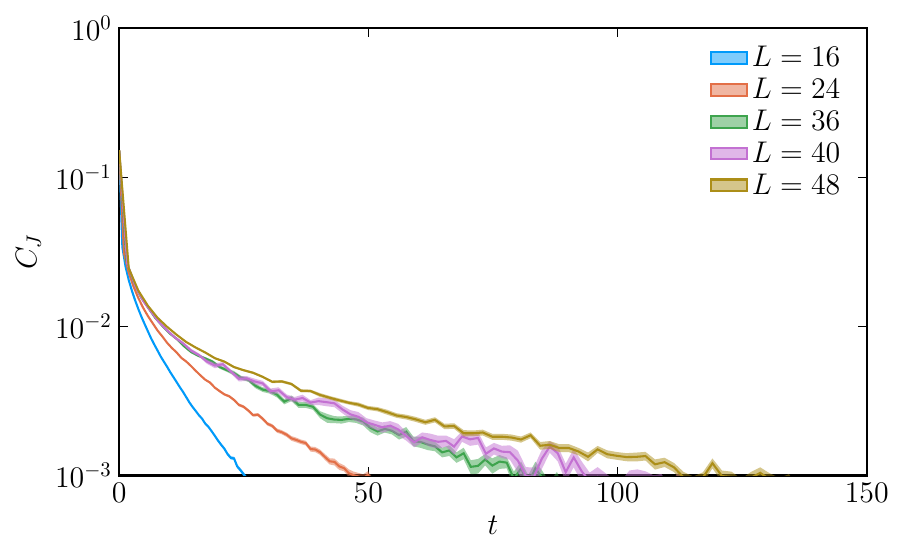}
\includegraphics[width=0.45\columnwidth]{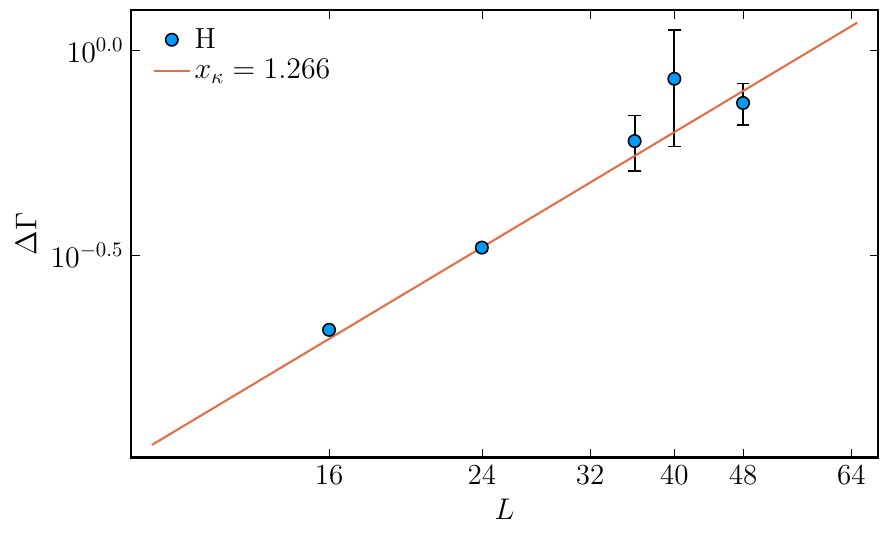}
\end{center}
\caption{\label{fig:kappa-3d-crit-H}
Same as Fig.~\ref{fig:kappa-3d-crit} in model H. Left panel
shows the Kubo integrand $C_\jmath(t)$ as a function of $t$, 
and the right panel presents $\Delta\kappa$ as a function
of $L$. Also shown is the best fit $\Delta\kappa\sim L^{x_\kappa}$ 
with $x_\kappa=1.266$.}
\end{figure}

\subsection{Critical transport in three dimensions}
\label{sec:num-3d-crit}

 Finally, we turn to critical transport in a three-dimensional 
fluid. Figs.~\ref{fig:kappa-3d-crit} and \ref{fig:kappa-3d-crit-H}
show the critical scaling of the thermal conductivity in model 
H and model H0. We observe that the Kubo integrand has a long-time tail, and
that the Kubo integral exhibits critical scaling. Compared to the calculation in
two dimensions the error bars are larger, and we have found it difficult to
extract a numerical value for the shear viscosity exponent $x_\eta$. We also
observe a bigger discrepancy between the values of $x_\kappa$ in model H0,
$x_\kappa=1.545$, and model H, $x_\kappa=1.266$. Without a result for $x_\eta$
we cannot check the first scaling relation. In model H the second scaling
relation is 
\begin{align}
z = 3.02 \hspace{1cm}
4-x_\kappa - \eta^* \approx 2.7\, ,
\end{align}
where we have again used the previously determined value of 
$z$ \cite{Chattopadhyay:2024bcv}. The discrepancy in model H0
is larger, $\Delta z\simeq 0.6$. Applying the arguments given 
in the previous section we estimate the systematic error in 
model H to be $\Delta x_\kappa\simeq 0.3$. Within this 
uncertainty, the values of $x_\kappa$ in model H0 and model H
are consistent with one another.

\section{Summary and outlook}
\label{sec:sum}

 In this work we have studied transport properties of fluids 
using a framework that takes into account thermal fluctuations
in the entropy per particle and momentum density. We focused
on Kubo relations as a tool to extract the shear viscosity 
and thermal conductivity. The Kubo relation is a well known
method for studying transport, and it is frequently used in 
connection with molecular dynamics or lattice quantum 
field theory simulations, but it has not been used in the 
context of stochastic fluid dynamics. The stochastic approach
is an effective field theory, and it includes UV divergences
not present in the microscopic theory that can be absorbed 
by local counterterms. The effective theory does faithfully
represent the IR behavior of the underlying microscopic theory. 

 We have verified that the numerical simulations indeed
reproduces the expected logarithmic IR divergence of the 
shear viscosity in a two-dimensional critical fluid. The 
calculations based on the Kubo integral also gives the linear 
UV divergence of the shear viscosity in three dimensions, 
but the coefficient of proportionality multiplying $a^{-1}$
does not agree with the result extracted from the decay 
of $\pi_i^T$ correlator. This is in contrast to the result
of a one-loop calculation in the continuum regularized by 
a three momentum cutoff. We conclude that within the lattice 
discretization used in the present work we cannot extract 
the relationship between the bare and renormalized viscosity
using the Kubo relation in eq.~(\ref{Kubo_eta_lat}).

  The Kubo relation can be used to study the critical 
behavior of transport coefficients in two and three dimensions. 
Our main findings are:

\begin{enumerate}
\item We observe critical scaling of both $\log(\eta)$
and $\log(\kappa)$ with $\log(L)$. The scaling holds in 
two as well as three dimensions, and for both model H and 
model H0. 

\item The critical divergence of $\kappa$ is significantly
stronger than that of $\eta$. This is consistent with the 
epsilon expansion. It is also not unexpected: The current 
$\vec\jmath=\phi\,\vec{\pi}^T$ couples directly to the order 
parameter whereas for the two terms in $\Pi_{ij}=\frac{1}{\rho}
\pi^T_i\pi^T_j+\nabla_i\phi\nabla_j\phi$, the first does not
directly couple to $\phi$, and the second is suppressed in 
the IR by powers of the wave number. 

\item Within the (sizable) uncertainties of our calculation 
the scaling relations in eq.~(\ref{x-kappa-eta},\ref{z-Gam})
are satisfied, and the critical exponents in model H0 and 
model H are consistent with one another.

\item The model H value for $x_\kappa$ is $x_\kappa\simeq 1.6
\pm 0.1$ in two dimensions, and $x_\kappa\simeq 1.25\pm 0.3$
in three dimensions. In $d=2$ we find $x_\eta\simeq 0.25\pm 
0.1$, and in $d=3$ we were unable to extract a numerical 
value for $x_\eta$. The value of $x_\eta$ in $d=2$ is consistent
with the epsilon expansion, eq.~(\ref{crit-eps}), but in 
(mild) tension with the functional renormalization group 
prediction $x_\eta=0$ \cite{Roth:2023wbp}. Note, however,
that for small values of $x$ it is hard to numerically 
distinguish $L^x$ and $x\log(L)$.

\end{enumerate}

 There are a number of important questions that we would 
like to address in the future. First, we would like to 
understand whether there is a discretization scheme that 
automatically respects the Kubo relation for the full shear 
viscosity, including the cutoff dependent part. Second, we
would like to explore methods for improving the numerical 
accuracy of $x_\kappa$ and $x_\eta$, including a determination
of $x_\eta$ in three dimensions. We can obviously collect 
more statistics, but the calculations presented here are 
already numerically quite demanding, in significant part 
because of the large value of the dynamic exponent $z\gsim 3$ 
of model H in three dimensions. One approach that might 
be helpful is to develop a better understanding of finite 
size scaling corrections to the Kubo integrand. Finally, 
we would like to understand the role of critical transport 
in simulations of realistic systems, such as the expanding 
quark gluon plasma produced in a relativistic heavy ion 
collision.

{\it Acknowledgments:}
This work is supported by the U.S. Department of Energy, Office 
of Science, Office of Nuclear Physics through the Contracts 
DE-FG02-03ER41260, DE-SC0024622 and DE-SC0020081. C.C. is supported by the Department of Space (DOS), Government of India. This work
used computing resources provided by the North Carolina State University High Performance Computing Services Core Facility 
(RRID:SCR-022168), as well as resources funded by the  Wesley
O.~Doggett endowment. 

{\it Data availability:}
Codes used to generate the data are available at 
\cite{Ott:2025a,Ott:2025b}.

\bibliography{bib}

\end{document}